\documentclass[aps,prb,floats,twocolumn]{revtex4}
\usepackage{epsfig}

\begin{document}         
\title{Magnetic phase diagram of the Kondo lattice model with quantum
  localized spins} 
\author{J. Kienert and W.~Nolting}
\affiliation{Lehrstuhl Festk{\"o}rpertheorie, Institut f{\"u}r Physik, Humboldt-Universit{\"a}t zu Berlin, 
  Newtonstr. 15, 12489 Berlin, Germany}
\begin{abstract}
The magnetic phase diagram of the ferromagnetic
Kondo lattice model is determined at T=0 in 1D, 2D, and 3D for
various magnitudes of the quantum mechanical localized spins ranging from
$S=\frac{1}{2}$ to classical spins ($S\rightarrow \infty$). We consider
the ferromagnetic phase, the paramagnetic phase, and the
ferromagnetic/antiferromagnetic phase separated regime. There is no
significant influence of the spin quantum number on the phase boundaries
except for the case $S=\frac{1}{2}$, where the model exhibits an
instability of the ferromagnetic phase with respect to spin disorder. 
Our results give support, at least as far as the low temperature
magnetic properties are concerned, to the classical treatment of the
$S=3/2-$spins in the intensively investigated manganites, for which the
ferromagnetic Kondo-lattice model is generally employed to account for
magnetism. 
\end{abstract}
\maketitle
\section{introduction}
The ferromagnetic Kondo lattice model (FKLM), also known as $s$-$d$
model or double exchange model, has drawn a lot of attention
over the past years in the field of magnetism and electronic
correlations. The model consists of Bloch electrons coupled to
localized spins sitting on the sites of a crystal lattice. 
For the case of strong (Hund) coupling and an energetically favored parallel
orientation of a localized moment and an electron, Zener proposed the
double exchange mechanism to 
explain ferromagnetism (FM) in manganites.\cite{Zen51} The gain in
kinetic energy of the conduction electrons favors a parallel
configuration of the localized spins. In the framework of a two-site
model, Anderson and Hasegawa showed that the hopping amplitude of the
electrons between sites $i$ and $j$ is proportional to
$cos(\theta_{ij}/2)$, where $\theta_{ij}$ is the angle between the
localized spins.\cite{AH55}   

A major field of application for the FKLM is linked to the phenomenon of 
colossal magnetoresistance\cite{Ram97} in the manganese compounds already
aimed at by Zener. Here,
the 5 Mn d-shells are split by the crystal field into three degenerate
$t_{2g}$-orbitals forming localized spins of $S=\frac{3}{2}$. They
interact via Hund's rule with itinerant electrons stemming from the
remaining two degenerate $e_{g}$-orbitals. With a Hund exchange interaction
mostly estimated to be several times the hopping
amplitude\cite{SPV96,MCS98,Dag03} the manganites belong to the rather
strongly coupled materials. Although there are other important aspects to take
into account when modelling the rich physics of the manganites, like
eletron-phonon-coupling and the second conduction band, the FKLM already in its simplest
single band version is crucial for understanding at least the magnetic
properties of this class of substances.\cite{Dag03}

A hot topic where the FKLM is used as a basic model are the diluted
magnetic semiconductors (DMS) 
with promising technical applications for microelectronics.\cite{Ohno99,ZFS04} These
materials consist of a (often III/V as, e.g., GaAs) semiconducting host
and substituted transition metal impurities (e.g. Mn)
occupying cation sites, the latters exhibiting ferromagnetism due to a
coupling of the localized cation spins mediated by a spin exchange
interaction with valence and impurity band holes. In the case of DMS, this
interaction is considered intermediate.\cite{HS05}

There have been various types of approaches to solving the many-body
problem of the FKLM in order to get a phase diagram. On the 
theoretical side, several treatments are based on Dynamical Mean Field
Theory (DMFT).\cite{Yun98,Dag98,NMK00,CMS01} In Ref. \onlinecite{Yin03} a
Schwinger-boson method is used and applied to 2D and 3D. Bosonization in 
1D is extensively discussed in Ref. \onlinecite{Gul04}. More recently an analytical 
continuum field theory in 2D was employed.\cite{PMTG05} Valuable information to
compare theoretical results with can be gathered from Monte Carlo
simulations.\cite{Yun98,Dag98,Ali01} The main feature of the magnetic phase 
diagram is the same in all these works: with increasing coupling strength 
ferromagnetism (FM) prevails for all charge carrier densities except for a more 
or less small region around half-filling where antiferromagnetism (AFM) or FM/AFM 
phase separation exists.  

Most approaches rely on the assumption that the local spins can be
treated classically. This 
simplification has been justified by checking the classical spin 
against quantum spin approaches, both giving similar
results.\cite{Yun98,Dag98} More recently, a phase diagram in 1D was
obtained by means of the Density-Matrix Renormalization Group (DMRG) 
yielding numerically exact results for a quantum spin $S=1/2$.\cite{GHAA04}
However, we do not know of any systematic,
quantitative analysis of the influence of the spin magnitude on the
magnetic properties of the FKLM.

In this work we present phase diagrams of the FKLM at T=0 for
several spin quantum numbers in 1D, 2D, and 3D. We use an equation of motion approach exploiting exactly
solvable limiting cases and exact relations among Green functions and
among their spectral moments. By evaluating the (free) energy at T=0
we can distinguish three different phases: paramagnetic (PM),
ferromagnetic, and ferromagnetic-antiferromagnetic phase separated
(PS). The latter two are the typical phases in the strong coupling region of the
phase diagram\cite{Yun98,Dag98,GHAA04,Yin03,PMTG05} which is the regime relevant for the manganites and on
which we want to focus in our work. It should be mentioned that in the
weak to intermediate coupling regime several other phases like canted,
spiral, or island have been found.\cite{GHAA04,Yin03,PMTG05}

One of the main results will
be that there are no major differences in the FM-PS phase boundaries for spin
quantum numbers $S>1/2$. The case $S=1/2$ is special: here we obtain an 
instability of the ferromagnetic against the paramagnetic phase. To
understand this behavior we discuss the spectral weight distribution of
the excitation spectrum and its modifications by a variation of the
quantum character of the localized spins.

The paper is organized as follows. After presenting our theoretical
approach in section II we discuss the phase diagrams of the 
FKLM in 1D, 2D, and 3D for different quantum numbers of the localized
spins in section III. A summary and an outlook on possible improvements
of our theory are given in section IV.

\section{Theory}

The Hamiltonian of the FKLM reads
\begin{equation}
  \label{Hklm}
  H= -t\sum_{\langle ij
    \rangle\sigma}c_{i\sigma}^{\dag}c_{j\sigma}-J\sum_{i}{\mathbf s}_{i}\cdot {\bf S}_{i}\;. 
\end{equation}
The first term describes Bloch electrons of spin $\sigma$ with a nearest
neighbor hopping integral $t$. $c_{i\sigma}^{({\dag})}$ annihilates (creates) an electron of spin $\sigma$ 
at lattice site $i$. The lattice is chosen to be simple cubic in our
case. $\mathbf s$ is the electron spin and $\bf S$ the localized spin operator, and both are
coupled through a Hund exchange $J$.
Using standard second quantization notation the interaction term can be rewritten
\begin{equation}
\label{Hklmquant}
H_{int}= -\frac{1}{2}J\sum_{i\sigma}z_{\sigma}S_{i}^{z}n_{i\sigma}+S_{i}^{-\sigma}c_{i\sigma}^{\dag}c_{i-\sigma}
\end{equation}
with $z_{\uparrow,\downarrow}=\pm 1$, $S_{i}^{z,+,-}$ are the z-component, raising and lowering operators for 
a localized spin at site $i$, and $n_{i\sigma}=c_{i\sigma}^{\dag}c_{i\sigma}$ is the occupation
number operator at site $i$.

The many-body problem of the above Hamiltonian is solved with the
knowledge of the one-electron Green function $G_{{\bf k}\sigma}(E)$, or, equivalently, the 
electronic self-energy $\Sigma_{{\bf k}\sigma}(E)$:
\begin{eqnarray}
\label{GF}
\nonumber
G_{ij\sigma}(E)&=&\langle\langle
c_{i\sigma};c^{\dag}_{j\sigma}\rangle\rangle_{E}=\frac{1}{N}\sum_{\bf
  k}G_{{\bf k}\sigma}(E)e^{i{\bf k}({\bf R}_i-{\bf R}_j)}\;,\\
G_{{\bf k}\sigma}(E)&=&\frac{\hbar}{E-\epsilon({\bf k})-\Sigma_{{\bf k}\sigma}(E)}\;,
\end{eqnarray}
with the Bloch dispersion $\epsilon({\bf k})$. One can then calculate the internal 
energy $U$ of the FKLM, being equivalent to the free energy at T=0,
for the ferromagnetic and the paramagnetic state. There is a simple
relation between the energy $U$ of the FKLM and the 
imaginary part of the corresponding one-particle Green function,
\begin{equation}
  \label{Utot}
  U=\langle H \rangle=\frac{1}{N\hbar}\sum_{i\sigma}\int_{-\infty}^{+\infty}dEf_{-}(E)ES_{ii\sigma}(E)\;,
\end{equation}
where $S_{ii\sigma}(E)=-\frac{1}{\pi}\Im G_{ii\sigma}(E)$ is the local
spectral density and $f_{-}$ denotes the Fermi function.\cite{remark1} 
 
Note that the existence of the ferromagnetic
state is supposed and not the result of a self-consistent calculation,
i.e. the magnetization is a parameter in our scheme.

The method we chose to solve the Hamiltonian (\ref{Hklm}) for the Green
function (\ref{GF}) is a moment conserving decoupling approach
(MCDA), which does {\em not} require the localized spins to be classical.
This theory has been applied before in model studies\cite{SN02} and to 
real substances\cite{ScN01,SEN04}. For a
detailed account of the decoupling procedure we refer the reader to
Ref.\cite{NRM97}. Here we summarize the main points of the
method and want to emphasize features which are important for the
following discussion of our results.

The starting point is the equation of motion for the Green function
(\ref{GF}). The generated higher Green functions read
\begin{eqnarray}
\label{I}
  I_{ik,j\sigma}(E)      &=&\langle\langle{S_{i}^{z}c_{k\sigma};c^{\dag}_{j\sigma}}\rangle\rangle_{E}\;,\\
\label{F}
  F_{ik,j\sigma}(E)      &=&\langle\langle{S_{i}^{-\sigma}c_{k-\sigma};c^{\dag}_{j\sigma}}\rangle\rangle_{E}\;.
\end{eqnarray}\\
$I_{ik,j\sigma}(E)$ is a Ising-like GF solely comprising the
z-components of the spins, whereas $F_{ik,j\sigma}(E)$ describes 
spin flip processes which are neglected when using classical localized
spins. After writing down the equations of motion of $I_{ik,j\sigma}(E)$
and $F_{ik,j\sigma}(E)$ the decoupling is 
performed. Of special importance for correlation effects is the
treatment of the local higher Green functions, namely 
\begin{eqnarray}
  \label{eq:F1}
  F^{(1)}_{ii,j\sigma}(E)&=&\langle\langle S_{i}^{-\sigma}S_{i}^{z}c_{i-\sigma};c^{\dag}_{j\sigma}\rangle\rangle_{E}\;,\\
  \label{eq:F2}
  F^{(2)}_{ii,j\sigma}(E)&=&\langle\langle S_{i}^{-\sigma}S_{i}^{\sigma}c_{i\sigma};c^{\dag}_{j\sigma}\rangle\rangle_{E}\;,\\
  \label{eq:F3}
  F^{(3)}_{ii,j\sigma}(E)&=&\langle\langle S_{i}^{-\sigma}n_{i\sigma}c_{i-\sigma};c^{\dag}_{j\sigma}\rangle\rangle_{E}\;,\\
  \label{eq:F4}
  F^{(4)}_{ii,j\sigma}(E)&=&\langle\langle S_{i}^{z}n_{i-\sigma}c_{i\sigma};c^{\dag}_{j\sigma}\rangle\rangle_{E}\;.
\end{eqnarray}
These functions are expressed in terms of the lower order GF (\ref{GF}), (\ref{I}), and (\ref{F}) 
with coefficients fitted by
the first two spectral moments of these GFs, respectively, representing a non-perturbative approximation for the
whole temperature range.\cite{NRM97} The choice of the "correct" lower order GF is guided by some non-trivial limiting 
cases which we summarize next. 

For an assumed complete
ferromagnetic polarization ($\langle S^{z}\rangle=S$) of the FKLM one obtains from 
the spectral representation of the Green functions: 
\begin{eqnarray}
\label{F1Sz_S}
   F^{(1)}_{ii,j\sigma}(E)&\stackrel{\langle S^z\rangle
     =S}{=}&\left(\left(S-\frac{1}{2}\right)+\frac{1}{2}z_{\sigma}\right)F_{ii,j\sigma}(E)\;,\\ 
\label{F2Sz_S}
   F^{(2)}_{ii,j\sigma}(E)&\stackrel{\langle S^z\rangle
     =S}{=}&SG_{ij\sigma}(E)-z_{\sigma}I_{ii,j\sigma}(E)\;.
\end{eqnarray}

For $S=\frac{1}{2}$, a case which is of particular importance to our
investigation due to its maximum amount of quantum fluctuations, the following 
relations hold at any temperature (i.e. at any $\langle S_{z} \rangle$):
\begin{eqnarray}
\label{F1S0_5}
   F^{(1)}_{ii,j\sigma}(E)&\stackrel{S=\frac{1}{2}}{=}&\frac{1}{2}z_{\sigma}F_{ii,j\sigma}(E)\;,\\
\label{F2S0_5}
   F^{(2)}_{ii,j\sigma}(E)&\stackrel{S=\frac{1}{2}}{=}&\frac{1}{2}G_{ij\sigma}(E)-z_{\sigma}I_{ii,j\sigma}(E)\;.
\end{eqnarray}
Furthermore in the case of a fully occupied conduction band:
\begin{eqnarray}
\label{F3_n2}
F^{(3)}_{ii,j\sigma}(E)&\stackrel{n=2}{=}&F_{ii,j\sigma}(E)\;,\\
\label{F4_n2}
F^{(4)}_{ii,j\sigma}(E)&\stackrel{n=2}{=}&I_{ii,j\sigma}(E)\;.
\end{eqnarray}

The above exact relations are used to motivate the following ansatz for the higher local GF:
\begin{eqnarray}
\label{F1Interpolation}
   F^{(1)}_{ii,j\sigma}(E)&=&\alpha_{1\sigma}G_{ij\sigma}(E)+\beta_{1\sigma}F_{ii,j\sigma}(E)\;,\\
\label{F2Interpolation}
   F^{(2)}_{ii,j\sigma}(E)&=&\alpha_{2\sigma}G_{ij\sigma}(E)+\beta_{2\sigma}I_{ii,j\sigma}(E)\;,\\
\label{F3Interpolation}
   F^{(3)}_{ii,j\sigma}(E)&=&\alpha_{3\sigma}G_{ij\sigma}(E)+\beta_{3\sigma}F_{ii,j\sigma}(E)\;,\\
\label{F4Interpolation}
   F^{(4)}_{ii,j\sigma}(E)&=&\alpha_{4\sigma}G_{ij\sigma}(E)+\beta_{4\sigma}I_{ii,j\sigma}(E)\;.
 \end{eqnarray}
The temperature dependent interpolation coefficients $\alpha_{i\sigma},~\beta_{i\sigma}~(i=1,..,4)$ 
depend on various correlation functions and are listed in Appendix A. It is easily verified that the 
approximations (\ref{F1Interpolation})-(\ref{F4Interpolation}) fulfill the exact limiting cases 
(\ref{F1Sz_S})-(\ref{F4_n2}). In addition our approach reproduces the limit of the ferromagnetically 
saturated semiconductor (T=0, band occupation n=0).\cite{Nol7} 

The resulting self-energy $\Sigma_{\sigma}(E)$ is local and depends on various 
expectation values of pure fermionic, mixed fermionic-spin, and pure
localized spin character:
\begin{eqnarray}
{{\Sigma}_{\sigma}}=F(\langle n_{\sigma}\rangle,
\langle
S_{}^{-\sigma}c_{\sigma}^{\dag}c_{-\sigma}\rangle,~\langle
  S^{z}n_{\sigma}\rangle,\\
\langle{S^{z}}\rangle,~\langle{(S^{z})^2}\rangle,~\langle{(S^{z})^3}\rangle,~\langle{S^{+}S^{-}}\rangle)\;.
\nonumber
\end{eqnarray}
The site indices have been dropped due to translational invariance.
Whereas the first two types can be calculated within the MCDA
the localized spin correlation functions are known for ferromagnetic saturation and 
in the paramagnetic state (see Appendix B). The many-body problem represented by (\ref{Hklm}) 
can thus be solved approximately for the Green function (\ref{GF}) in a self-consistent 
manner.\cite{NRM97} We emphasize that the quantum 
mechanical character of the localized spins is fully retained in our approach. Furthermore 
there is no restriction to the parameter range within which our method can be applied.

In order to determine the phase boundary between the ferromagnetic and
the ferromagnetic-antiferromagnetic phase separated region we first have
to solve the Hamiltonian (\ref{Hklm}) for an antiferromagnetic
configuration. Using the standard sublattice decomposition for bipartite lattices and 
neglecting the off-diagonal elements of the self-energy matrix\cite{NMR96}
one obtains the following Green function for sublattice A:
\begin{eqnarray}
G^{A}_{{\bf k}\sigma}(E)&=&\frac{\hbar}{E-\epsilon'({\bf
    k})-\Sigma^{A}_{{\bf k}\sigma}(E)-\frac{|t({\bf
      k})|^2}{E-\epsilon'({\bf k})-\Sigma^{B}_{{\bf k}\sigma}(E)}}\;, 
\end{eqnarray}
with the diagonal elements of the self-energy matrix $\Sigma^{A}_{{\bf
    k}\sigma}=\Sigma^{B}_{{\bf k}-\sigma}$, the sublattice dispersion
$\epsilon'({\bf k})$ and the inter-sublattice dispersion $t({\bf
  k})$. The approximate solution for the self-energy presented above for the
translationally invariant case can be obtained analogously for the
antiferromagnetic case. The energy of the antiferromagnetic phase 
can be evaluated using (\ref{Utot}) by simply replacing $S_{ii\sigma}(E)$ 
by $S^{A}_{ii\sigma}(E)$. Averaging over the sublattices is not necessary 
due to symmetry reasons, i.e. the summation over the sublattices is absorbed into 
the spin summation.

In this work we restrict our considerations concerning AFM to $G$-type
antiferromagnetism, i.e. the spins of all nearest neighbors of a given
lattice site belong to the other sublattice. This kind of antiferromagnetism
is typical in the strong coupling regime at and near half-filling  
because it allows for a maximum kinetic energy gain by virtual hopping processes, unlike a FM configuration 
which forbids these by Pauli's Principle. We assume the ground state has N$\rm\acute{e}$el
structure, i.e. we set the two magnetic sublattices to be saturated, $\langle S_{z}^{A}\rangle = S = -\langle 
S_{z}^{B}\rangle$. Furthermore we do not take into account possible
canted AFM configurations. 
 
We used the method proposed in Ref. \onlinecite{Yin03} to determine the FM/PS 
phase boundary.\cite{remark2} This criterium for electronic phase separation is based 
on the separation into AFM regions with one electron at each site 
and FM domains with an occupation $n_c$, a picture suggested by numerical results in Ref. 
\onlinecite{Dag98}. On increasing $n$ the AFM part grows until it occupies the 
whole system at half-filling. The total energy can be written as  
\begin{eqnarray}
  \label{U_tot_PS}
  U_{tot}(n,v) = (1-v)U_{AFM} + vU_{FM}\left(1-\frac{1-n}{v}\right)
\end{eqnarray}
where $U_{tot}$ is the total energy per site, $U_{FM}$ is the FM energy 
per site and its argument is the particle density in the FM regions ($n$ is the total 
particle density), $U_{AFM}$ is the AFM 
energy per site, and $v=V_{FM}/V_{tot}$ is the FM volume fraction of the total system size. 
Minimizing the total energy with respect to $v$ yields the following condition for the critical 
electron density $n_c$ at which electronic phase separation sets in (i.e. $v=1$):
\begin{equation}
  \label{E_minimum}
  U_{AFM}=U_{FM}(n_c)+(1-n_c)U'_{FM}(n_c)\;.
\end{equation}
$U'_{FM}$ is the derivative of $U_{FM}$ with respect to the particle density.
Note that we consider the electron density and not the hole
density in the FKLM so that the corresponding formula in
Ref. \onlinecite{Yin03} is modified accordingly. 

If one varies the chemical potential $\mu$ continuously
the electron densities at which phase separation is present correspond
to band occupations that cannot be stabilized as, e.g., demonstrated in
the MC simulations in Ref. \onlinecite{Dag98}. Given the jump in the electron
density on varying $\mu$ this kind of transition appears to be
first-order. However, {\em enforcing} any value of $n$ as in our case
and having in mind the picture of AFM regions gradually taking over the whole
system the transition from FM to AFM rather appears to be continuous.

\section{Results and Discussion}

We computed phase diagrams for different spin quantum numbers $S$ 
for a simple cubic lattice, a square lattice, and a 1D chain at zero temperature. 
To simulate a classical spin, i.e. a spin that can be oriented in any direction in space, 
we used a spin quantum number $S=10$. In order to obtain a FM-PM-PS phase diagram we 
evaluated the total energy at T=0 for the saturated ferromagnetic, for the
paramagnetic, and for the antiferromagnetic ($n=1$) spin configuration
of the core spins, respectively, as a function of the occupation number
$n$ and for several values of the Hund coupling $J$. To compare the
results for different $S$ we take the proper scaling $\propto JS$ of
the interaction into account and normalize the localized spins. Thus in
the following we consider localized spins ${\bf S}/S$ coupled to
itinerant electrons by $JS$.
\begin{figure}[t]
\epsfig{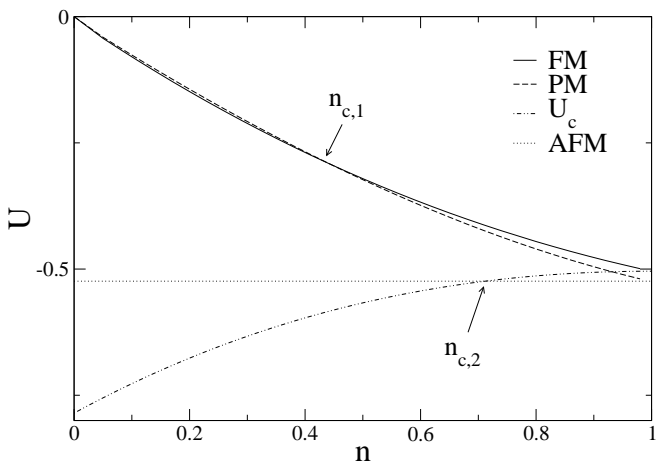}
\caption{Total energy (in eV) for ferromagnetic, paramagnetic, and
  antiferromagnetic configuration for the square lattice. The AFM energy
  is evaluated at $n=1$. $U_c$ is computed according to the right hand
  side of Eq.~(\ref{E_minimum}). Parameters: $S=1/2,~JS/t=12$.}   
\label{U1}
\end{figure} 

Before starting the discussion of our results we should add two
remarks. First, it is well known that 1D systems exhibit some
pecularities; for example, an integer spin nearest-neighbor Heisenberg
chain has a gap in its excitation spectrum ({\em Haldane gap}). Moreover,
non-local correlations are important, whereas our approach is based on a
local self-energy. However, our approximate theory is applicable for
any finite dimension and thus we consider it worthwhile to present results for
$D=1$, too.

Secondly, we have to address the issue of anisotropy. The Mermin-Wagner
theorem\cite{MW66} forbids spontaneous symmetry breaking in $D<3$  
for an isotropic Hamiltonian like (\ref{Hklm}) at finite temperatures. In order for 
our results to be relevant at small temperatures, too, we have to add an 
(infinitesimally) small anisotropy term, e.g. a single-ion anisotropy taking 
spin orbit coupling into account. Being orders of magnitude smaller than the leading 
energy scale in our system, the Hund coupling $J$, it will not alter the phase 
boundaries visibly. There is another benefit of adding an anisotropy to (\ref{Hklm}). 
On decreasing $S$ the assumption of a N$\rm\acute{e}$el state for the antiferromagnetic 
phase becomes questionable due to quantum fluctuations. In 1D this approximation 
even breaks down completely. These fluctuations are suppressed by
anisotropy.\cite{Faz99}

\begin{figure}[t]
\epsfig{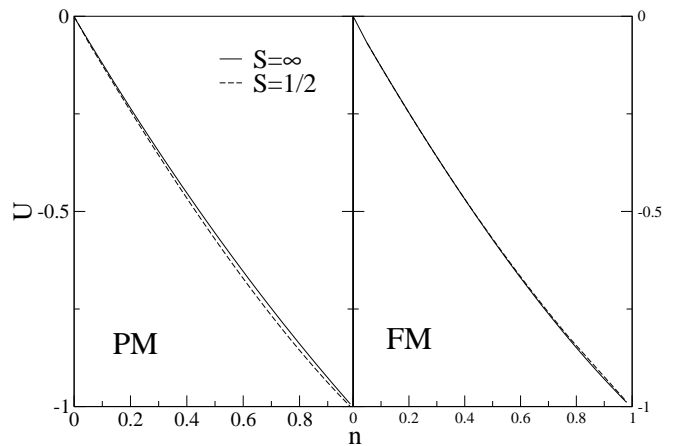}
\caption{PM (left) and FM (right) total energy (eV) of the square lattice for
  quantum $S=1/2$ and classical localized spins. Parameters: $JS/t=24$.}
\label{U2}
\end{figure}
\begin{figure}[t]
\epsfig{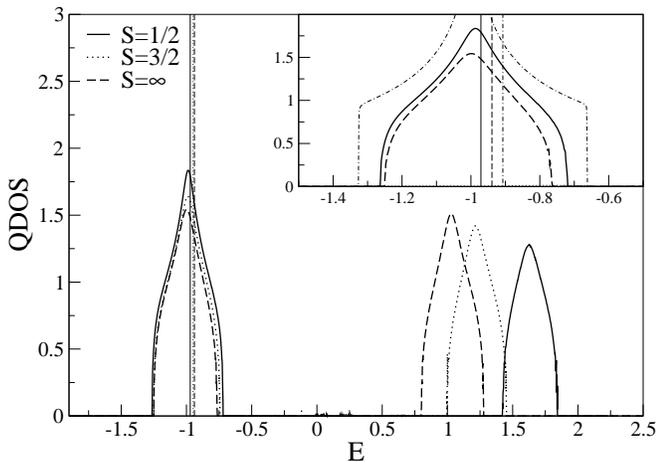}
\caption{Paramagnetic quasiparticle density of states of the square lattice. Parameters: JS/t=24, n=0.7. The inset 
shows the low energy band of the spectrum on a smaller scale. Vertical lines indicate the
chemical potential. The dash-dotted line in the inset is the Bloch-shaped FM $\uparrow$-QDOS ($S=1/2$). 
The FM $\downarrow$-QDOS in this energy range practically vanishes (dotted line) indicating complete spin 
polarization.}  
\label{qdos_PM}
\end{figure} 
\begin{figure}[b]
\epsfig{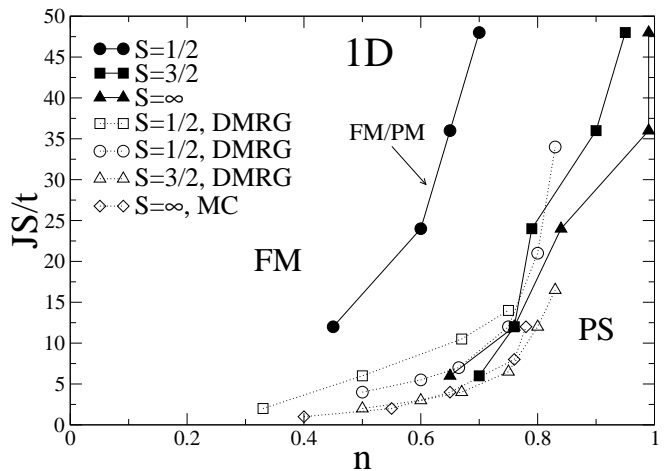}
\caption{Phase diagram in $D=1$ for
  $S=\frac{1}{2},\frac{3}{2}$ and classical spin. Filled symbols refer to our 
results and indicate the FM/PS transition ($S=3/2, \infty$) and the FM/PM 
transition ($S=1/2$). Open squares from
  Ref. \onlinecite{GHAA04} mark the transition from FM to a spiral phase; 
in this work an island phase between $n\approx 0.2$ and $n\approx 0.8$ and up to $JS/t\approx 8$, and PS 
for $n\geq 0.8$ and $JS/t\geq 10$ were also found. All other open symbols are taken 
from Ref. \onlinecite{Dag98}: circles ($S=1/2$) and triangles ($S=3/2$) indicate the 
boundary between FM and incommensurate correlations (IC), for $n\geq0.8$ and above $JS/t\approx 12~(S=1/2)$ and
$JS/t\approx 6~(S=3/2)$ PS was found, too; diamonds ($S=\infty$) mark transition from FM to PS ($JS/t\geq4$) 
and from FM to IC at weaker coupling. Lines are guides to the eye.}   
\label{pd_1D}
\end{figure} 

\begin{figure}[t]
\epsfig{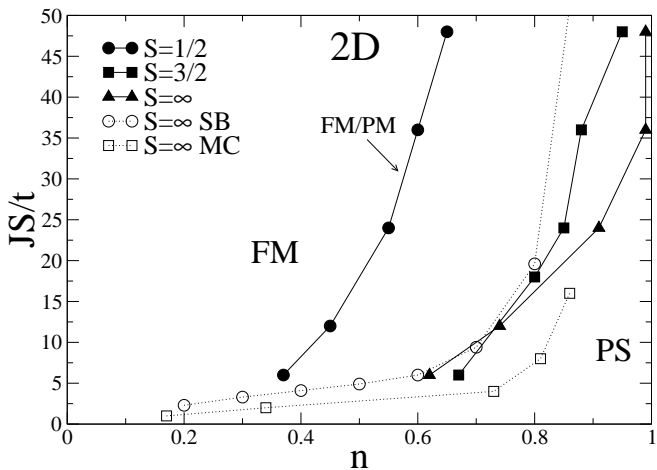}
\caption{Phase diagram in $D=2$ for
  $S=\frac{1}{2},\frac{3}{2}$ and classical spin. Open circles from 
Ref. \onlinecite{Yin03} ($S=\infty$) indicate boundary between FM and PS above $JS/t\approx 7$
 and between FM and a spiral phase at weaker coupling. SB stands for Schwinger-boson. 
Open squares from Ref. \onlinecite{Yun98} ($S=\infty$) mark transition from 
FM to PS above $JS/t\approx 4$ and from FM to IC at weaker coupling.
Lines are guides to the eye.} 
\label{pd_2D}
\end{figure} 

Fig. \ref{U1} shows the total energy of the FM and PM phases as a
function of the band occupation $n$ and of the AFM phase at $n=1$. The
result of the right hand side of Eq. (\ref{E_minimum}) is also
plotted. The spin is $S=1/2$ corresponding to a maximum of quantum
fluctuations. The paramagnetic ground state at $n_{c,1}=0.45$ emerges 
well before the criterium (\ref{E_minimum}) for PS is fulfilled at
$n_{c,2}=0.72$, indicating an instability of ferromagnetism against spin
disorder. 
In our calculations we did not find a second transition from PM to PS for $S=1/2$ 
at $n>n_{c,1}$. On the other hand for $S\ge 1$ and the values of $J$ we considered 
($JS/t\ge6$) we find that the critical value of $n$ for the onset of phase separation is 
always lower than the electron density where FM becomes unstable against PM, i.e. 
$n_{c,2}<n_{c,1}$. 

To further analyse the FM-PM transition for $S=1/2$ we show in
Fig. \ref{U2} more results for the total energy and compare them to
calculations based on classical localized spins. Whereas the PM energy
is lower for the quantum spin over the whole range of electron densities
the FM energies are practically the same for both spins. 

\begin{figure}[t]
\epsfig{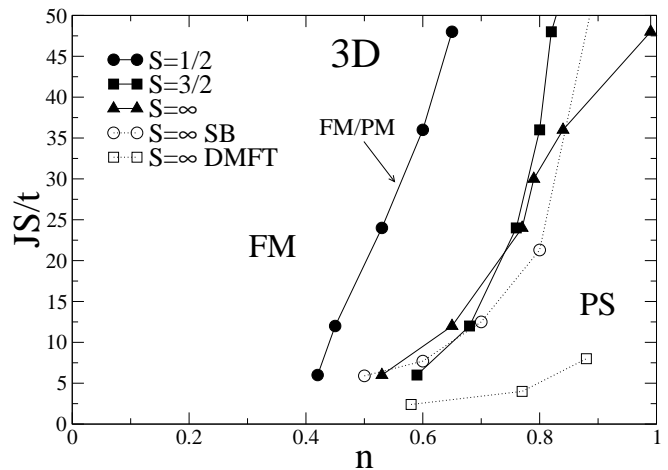}
\caption{Phase diagram in $D=3$ for
  $S=\frac{1}{2},\frac{3}{2}$ and classical spin. Open circles from 
Ref. \onlinecite{Yin03} indicate boundary between FM and PS above $JS/t\approx 7$ 
and between FM and spiral phase at weaker coupling.
Open squares from Ref. \onlinecite{CMS01} (semicircular density of states) 
mark transition from FM to PS. Lines are guides to the eye.}
\label{pd_3D}
\end{figure} 

This can be related to the quasiparticle excitation spectrum. As can 
be seen by (\ref{Utot}) the total energy of the FKLM is determined in
complete formal analogy to the free electron case,
i.e. by the (quasiparticle) density of states. Our findings suggest the
following picture: In the FM phase there is a parallel alignment of the conduction electrons with 
respect to the localized spins. Given a saturated FM spin background 
it is not possible for an itinerant electron to flip its spin by spin exchange. 
There is no significant occupation in the $\downarrow$-band of the
spectrum for any $S$. This is, at least for high Hund coupling,
consistent with the excitation spectra we obtained (see FM QDOS in the inset of Fig. \ref{qdos_PM}). 
For the $\uparrow$-electrons the localized moments act 
as an effective field and their quasiparticle density of states (QDOS) is
merely the Bloch density of states shifted by $-\frac{1}{2}JS$. Within
this picture the scaling by $JS$ is expected to transfer directly to the
total energy which is indeed the case as can be seen by the practically identical curves in
Fig. \ref{U2} (right). In other words, the scaling of the FM energy with
$JS$ expresses the fact that spin waves are frozen. 

The situation is different in the paramagnetic case. Now there are (energetically 
unfavored) states for $\uparrow$- and $\downarrow$-electrons with an antiparallel 
orientation to the localized spins. Whereas the lower
energy band in Fig. \ref{qdos_PM} has a "parallel character" of localized and itinerant
magnetic moment the upper band corresponds to an antiparallel
orientation. As there are more eigenstates with a parallel spin-spin
alignment one expects larger spectral weight of the corresponding peaks
in the excitation spectrum. The lower the magnitude of the localized
spin the higher this difference becomes: the 
parallel case "outweighs" the antiparallel case most dominantly for
$S=1/2$ ("triplet" vs. "singlet"). This can immediately be verified in
the zero bandwidth limit.\cite{Nol7} There is indeed a higher spectral
weight for low spin quantum numbers as can be seen in
Fig. \ref{qdos_PM}. On the other hand increasing $S$ results in an
equal distribution of spectral weight for both bands in the classical
limit.\cite{remark3} Thus, the paramagnetic state for lower magnitudes 
of the localized spins has lower total energy, as is observed in
Fig. \ref{U2} (left). 

Let us now proceed to the discussion of the magnetic phase diagrams we obtained
in the spatial dimensions $D=$1,2, and 3. Fig. \ref{pd_1D} shows the phase
diagram in $D=1$. The phase boundaries indicate the transition from FM
to the phase separated FM-AFM regime except for the case $S=1/2$ which
gives a FM-PM transition. For comparison we added results by other
authors obtained by numerical methods. Our findings give the same overall
picture as earlier works with an increasing FM region as the
Hund coupling becomes stronger. However we find that the $S=1/2$-FKLM has a
significantly reduced ferromagnetic stability due to its "early" transition to a
PM state compared to higher localized spin quantum numbers. A second
important feature is that we do not see any major 
differences in our phase boundaries for $S>1/2$.

Our results compare well with the MC numerical phase diagram (classical
spins) and with the DMRG results ($S=3/2$). There is no considerable
variation of the phase boundary for $S=3/2$ and $S=\infty$
either.\cite{Yun98} The deviations are largest for $S=1/2$: here our
theory appears to underestimate the FM region. However, the authors of
Ref. \onlinecite{GHAA04} attribute the fact that their FM region is reduced as
compared to Ref. \onlinecite{Dag98} to the higher number of lattice sites they
include in their computation. Hence it would be desirable to have more DMRG 
results with larger system sizes to see how the phase boundaries change. We did not 
find a second transition from PM to PS for $S=1/2$, neither did we take
spiral or incommensurate correlations into account as was done in Ref. 
\onlinecite{GHAA04} and Ref. \onlinecite{Dag98}. However we can state that 
the reduction of the {\em maximal} region of FM for $S=1/2$ with respect
to higher spin quantum numbers is consistent with what we learn from the
results obtained by other approaches. 

Fig. \ref{pd_2D} and \ref{pd_3D} show the phase diagrams for 2D and 3D,
respectively.\footnote{We point out that the 2D results taken from Ref. \onlinecite{Yun98} 
into Ref. \onlinecite{Yin03} for comparison are larger than the originally published 
data by a factor 2.} They give essentially the same 
picture as in 1D. Again we observe an unstability of the FM phase against PM for $S=1/2$ only,
reducing the region of FM stability as compared with higher $S$. There
is no significant change of the phase boundary for $S>1/2$ apart from some
enhancement of FM for classical core spins and larger Hund couplings in
all dimensions. The $S=\infty$ results in 2D are in accordance with MC 
simulations in Ref. \onlinecite{Ali01} which yield FM for the full range of band filling 
at large $J$. We note a slight enlargement of the PS region in
3D for all $S$. However we do not want to emphasize the quantitative
differences too much. As was already pointed
out in Ref. \onlinecite{CMS01} the small differences in the energies of the
different phases lead to uncertainties in the exact location of the
phase boundaries. In our case we estimate these error bars to be about
10\% with respect to the corresponding electron density $n$. For the same reason we 
are careful not to put too much significance into the $D$-dependence of the crossing 
points of our $S=3/2$ and classical $S$ phase boundaries. However it is interesting to 
note that there is a crossing in all dimensions.

As before our findings agree well with those published by other authors who used
different methods. There is one exception: compared to the other results
the DMFT phase boundaries (however, for a $D=\infty$-Bethe lattice)
from Ref. \onlinecite{CMS01} seem to overestimate FM considerably.

We conclude that the magnetic properties of the FKLM at strong coupling and as far as 
the magnetic phases we investigated are concerned are rather insensitive to variations 
of the spatial dimension, at least at T=0. This falls in line with other results obtained 
using classical localized spins.\cite{Dag98}

\section{Summary and Outlook}

We have presented magnetic phase diagrams for the ferromagnetic
Kondo lattice model in $D=1,2,$ and $3$. Our method is based on an
equation of motion decoupling procedure fulfilling exact relations among
Green functions and among their spectral moments. It does not require
the assumption of classical spins. To determine the phase boundaries we
computed and compared the total energy of the different phases at zero
temperature.

There are three main results. First the case $S=1/2$ appears to be special exhibiting a
reduced region of ferromagnetism in the $J$-$n$-plane due to
an instability of FM against spin disorder. Increasing the electron density
this transition always takes place before FM/AFM phase separation can
occur. Secondly the phase boundaries for $S>1/2$ appear to be quite
robust with respect to changes of the magnitude of the localized
spin. This supports the widespread usage of classical localized spins
in the treatment of the FKLM. Finally, these two features are recovered
and quantitatively similar in the phase diagrams in all dimensions we
investigated, namely 1D, 2D, and 3D.

To our knowledge there are no numerical phase diagram results
in 2D and 3D with quantum localized spins as this is numerically a quite
demanding task. It would be interesting to explore if the same trends
as in 1D hold for $S=1/2$ using numerically exact methods like
DMRG. It should also be a worthwhile task to examine the influence of
the spin magnitude at higher temperatures up to the Curie temperature
(in the vicinity of which colossal magnetoresistance occurs in the
manganites). Phase boundaries may of course change to a certain extent with the
crystal lattice structure, i.e. Bloch density of states. More changes can be expected
when including a finite next-nearest neighbor hopping integral. Finally
we focussed on the phases thought to be relevant for the intermediate to
strong-coupling regime and left out other phases that come into play in
the weak-coupling case. These issues are left for further investigation.

\section*{Appendix A: Interpolation coefficients}

Exploiting spectral moment relations leads to the following coefficients in the approximations 
(\ref{F1Interpolation})-(\ref{F4Interpolation}) for the higher order local Green functions:

\begin{eqnarray}
  \label{eq:interpolkoeff1}
  \alpha_{1\sigma}&=&0\\
  \label{eq:interpolkoeff2}
  \beta_{1\sigma}&=&\frac{K_{1\sigma}+4\Delta_{-\sigma}-3z_{\sigma}\mu_{-\sigma}-\eta_{\sigma}}{\langle
      S^{-\sigma}S^{\sigma}\rangle+2z_{\sigma}\Delta_{-\sigma}-\gamma_{\sigma}}\\\nonumber\\
  \label{eq:interpolkoeff3}
    \alpha_{2\sigma}&=&\langle
    S^{-\sigma}S^{\sigma}\rangle-\beta_{2\sigma}\langle S^{z}\rangle\\
   \label{eq:interpolkoeff4}
     \beta_{2\sigma}&=&\frac{K_{2\sigma}+2\eta_{\sigma}}{\langle
       (S^{z})^{2}\rangle-\langle
       S^{z}\rangle^{2}-\gamma_{\sigma}}\\\nonumber\\
   \label{eq:interpolkoeff5}
     \alpha_{3\sigma}&=&-\gamma_{\sigma}\\
   \label{eq:interpolkoeff6}
     \beta_{3\sigma}&=&\frac{\mu_{\sigma}-z_{\sigma}\eta_{\sigma}+2z_{\sigma}\vartheta+z_{\sigma}\gamma_{\sigma}\langle
       S^{z}\rangle}{\langle
       S^{-\sigma}S^{\sigma}\rangle+2z_{\sigma}\Delta_{-\sigma}-\gamma_{\sigma}}\\\nonumber\\
   \label{eq:interpolkoeff7}
     \alpha_{4\sigma}&=&\Delta_{-\sigma}+\beta_{4\sigma}\langle
     S^{z}\rangle\\
   \label{eq:interpolkoeff8}
     \beta_{4\sigma}&=&\frac{z_{\sigma}K_{3\sigma}-\mu_{-\sigma}-z_{\sigma}\eta_{\sigma}}{\langle
       (S^{z})^{2}\rangle-\langle
       S^{z}\rangle^{2}-\gamma_{\sigma}} 
  \end{eqnarray}\\
with the abbreviations:
  \begin{eqnarray}
K_{1\sigma}&=&3z_{\sigma}\langle
      S^{\sigma}S^{-\sigma}\rangle+(S(S+1)-4)\langle
      S^{z}\rangle +z_{\sigma}\langle
      (S^{z})^{2}\rangle \nonumber\\
&&-2z_{\sigma}S(S+1)(1-\langle
      n_{-\sigma}\rangle)
-\langle(S^{z})^{3}\rangle\\ 
K_{2\sigma}&=&\left(S(S+1)-\langle
      S^{-\sigma}S^{\sigma}\rangle\right)\langle
      S^{z}\rangle \nonumber\\
 &&-z_{\sigma}\langle
      (S^{z})^{2}\rangle-\langle(S^{z})^{3}\rangle\\
K_{3\sigma}&=&z_{\sigma}S(S+1)\langle
      n_{-\sigma}\rangle+\Delta_{-\sigma}(1-z_{\sigma}\langle
      S^{z}\rangle)
\end{eqnarray}
The mixed expectation values
\begin{eqnarray}
  \label{gamma}
  \gamma_{\sigma}&=&\langle
  S^{-\sigma}c^{\dag}_{\sigma}c_{-\sigma}\rangle\\
  \label{Delta}
  \Delta_{\sigma}&=&\langle S^{z}n_{\sigma}\rangle\\
  \mu_{\sigma}&=&\langle S^{-\sigma}S^{\sigma}n_{\sigma}\rangle\\
  \eta_{\sigma}&=&\langle
  S^{-\sigma}S^{z}c_{\sigma}^{\dag}c_{-\sigma}\rangle\\
  \vartheta&=&\langle S^{z}n_{\sigma}n_{-\sigma}\rangle
\end{eqnarray}
can all be evaluated with the corresponding Green functions using the spectral theorem.

\section*{Appendix B: Localized spin expectation values}

It holds for ferromagnetic saturation:
\begin{eqnarray}
  \label{eq:spinerwartungswerteferro}
  \langle{(S^{z})^2}\rangle             &=&S^2\\
  \langle{(S^z)^3}\rangle               &=&S^3\\
  \langle{S^{-\sigma}S^{\sigma}}\rangle &=&S(1-z_{\sigma})
\end{eqnarray} 

and for the paramagnetic phase:
\begin{eqnarray}
  \label{eq:spinerwartungswertepara}
  \langle{(S^{z})^2}\rangle             &=&\frac{1}{3}S(S+1)\\
  \langle{(S^z)^3}\rangle               &=&0\\
  \langle{S^{-\sigma}S^{\sigma}}\rangle &=&\frac{2}{3}S(S+1)
\end{eqnarray}


\begin{thebibliography}{14}
\bibitem{remark1} 
In the FKLM the total energy can be related to the local one-electron
Green function. This is due to the bilinearity of the
Hamiltonian in the fermionic operators. In other models, e.g. the Hubbard model, the 
corresponding formula is slightly more complicated and contains non-local terms.
\bibitem{remark2} 
By PS we always denote the coexistence of FM and AFM 
phases in the system.
\bibitem{remark3}
The different centers of gravity of the upper band are
due to a different scaling behavior. In the zero bandwidth
limit\cite{Nol7} the "antiparallel" excitation level scales as $J(S+1)$
rather than $JS$. In the framework of our theory the upper energy band
can be interpreted in fact as a mixture of two bands, where one can be
identified with the already mentioned parallel alignment of an electron and a
localized magnetic moment and the other is linked to the break-up of a
parallel coupling by a second conduction electron.\cite{JK1}
\end{thebibliography}
\end{document}